\documentclass[twocolumn,pra,superscriptaddress,aps,amsmath,floatfix]{revtex4-1}

\usepackage{graphics}
\usepackage{graphicx}
\usepackage{epstopdf}
\usepackage{dcolumn}
\usepackage{bm}
\usepackage{longtable}
\usepackage{epsfig}
\usepackage{times}
\usepackage{url}
\usepackage{color}

\begin{document}
\title{Setting a disordered password on a photonic memory}


\author{Shih-Wei  \surname{Su}}
\affiliation{Department of Physics and Graduate Institute of Photonics, National Changhua University of Education, Changhua 50058, Taiwan}
\author{Shih-Chuan \surname{Gou}}
\affiliation{Department of Physics and Graduate Institute of Photonics, National Changhua University of Education, Changhua 50058, Taiwan}
\author{Lock Yue  \surname{Chew}}
\affiliation{Division of Physics and Applied Physics, School of Physical and Mathematical Sciences, Nanyang Technological University, 21 Nanyang Link, Singapore 637371, Singapore}
\author{Yu-Yen  \surname{Chang}}
\affiliation{CEA Saclay, Service d'Astrophysique, 91191 Gif-sur-Yvette Cedex, France}
\affiliation{Academia Sinica Institute of Astronomy and Astrophysics, Taipei 10617, Taiwan}
\author{Ite A.  \surname{Yu}}
\affiliation{Department of Physics and Frontier Research Center on Fundamental and Applied Sciences of Matters, National Tsing Hua University, Hsinchu 30013, Taiwan}
\author{Alexey \surname{Kalachev}}
\affiliation{Zavoisky Physical-Technical Institute of the Russian Academy of Sciences, Sibirsky Trakt 10/7, Kazan, 420029, Russia}
\author{Wen-Te \surname{Liao}}
\email{wente.liao@g.ncu.edu.tw}
\affiliation{Department of Physics, National Central University, Taoyuan City 32001, Taiwan}
\date{\today}
\begin{abstract}
Encryption is a vital tool of information technology  protecting our data in the world with ubiquitous computers.
While photons are regarded as ideal information carriers, it is a must to implement such data protection on all-optical storage. However, the intrinsic risk of data breaches in existing schemes of photonic memory was never addressed.
We theoretically demonstrate the first protocol using spatially disordered laser fields to encrypt data stored on an optical memory, namely, encrypted photonic memory. Compare with a digital key, a continuous disorder encrypts stored light pulses with a rather long key length against brute-force attacks.
To address the broadband storage, we also investigate a novel scheme of disordered echo memory with a high fidelity approaching unity. Our results pave novel ways to encrypt different schemes of photonic memory  based on quantum optics and raise the security level of photonic information technology.
\end{abstract}

\maketitle

\begin{figure*}[t]
\vspace{-0.4cm}
  \includegraphics[width=1\textwidth]{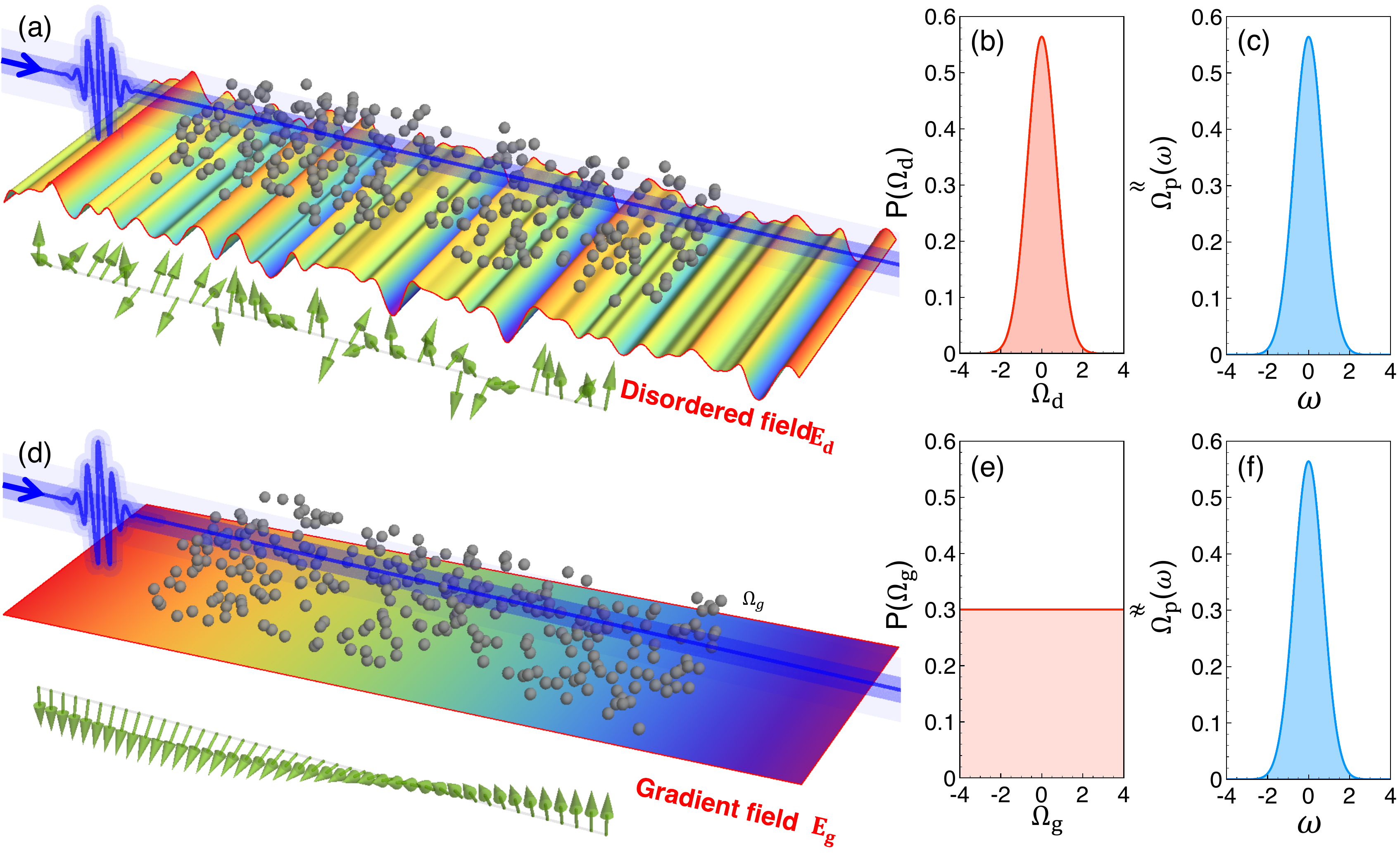}
  \caption{\label{fig1}
(Color online). 
(a) Disordered echo memory is composed of an atomic ensemble interacting with a spatially disordered field (coloured surface) and a probe pulse (blue line) to be stored.
The green randomly oriented arrows demonstrate the quantum phase evolution of memory atoms at various sections of the disorder. 
(b) The  probability distribution of the Rabi frequency of a disordered field should cover
(c,f) Fourier spectrum of the incident probe field.
(d) Gradient echo memory is made of an atomic ensemble interacting with a linearly gradient field which introduces a spatially ordered phase modulation. 
(e) The equal probability distribution of the linearly gradient field strength is distinct from (f).
  }
\end{figure*}
\begin{figure*}[t]
\vspace{-0.4cm}
  \includegraphics[width=0.9\textwidth]{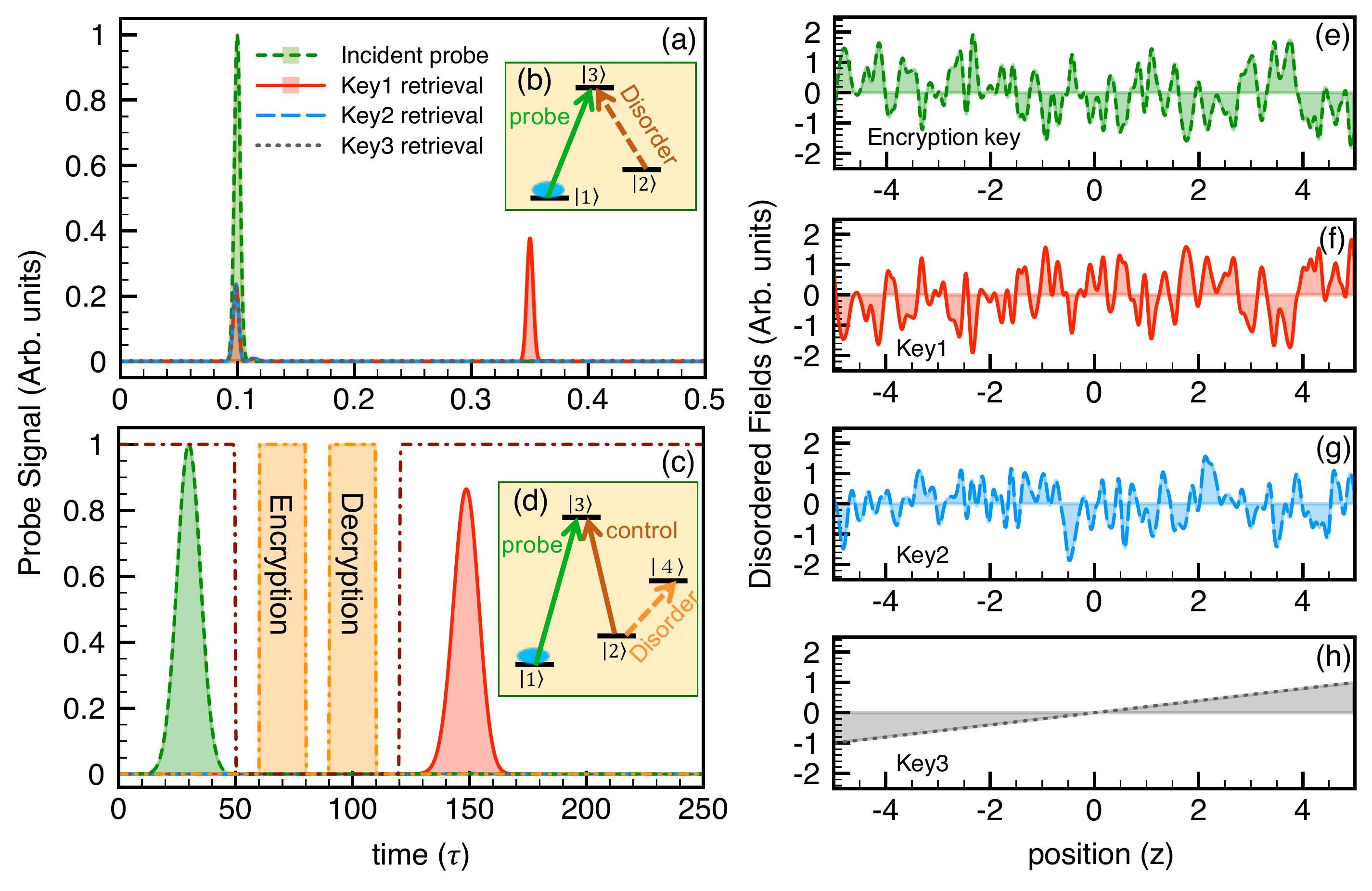}
  \caption{\label{fig2}
(Color online). 
Panel (a) illustrates broadband disordered echo memory based on (b) three-level $\Lambda$-type scheme, and 
(c) depicts narrowband EIT memory based on (d) four-level $N$-type system.
Incident probe pulses (green dashed lines) driving transition $\vert 1\rangle\leftrightarrow \vert 3\rangle$ are encrypted by the disorder in (e),
and  decrypted  by different keys in (f-h). 
The red solid lines depict the only successful retrieval with the correct key in (f). The blue long-dashed lines and gray dotted lines are the attempted retrievals by  two wrong keys depicted respectively in (g) and (h). 
In (a)
the access and encryption of a broadband probe are accomplished by
the disorder, inverted at $t= 0.22\tau$ for decryption, driving transition $\vert 2\rangle\leftrightarrow \vert 3\rangle$.
In (c) the brown dashed dotted line illustrates the temporal sequence of control field for EIT storage. Two disordered pulses (orange dashed dotted dotted line) driving transition $\vert 2\rangle\leftrightarrow \vert 4\rangle$ are applied, in turn, to encrypt and decrypt quantum coherence.
}
\end{figure*}
%
\begin{figure}[b]
\vspace{-0.4cm}
  \includegraphics[width=0.45\textwidth]{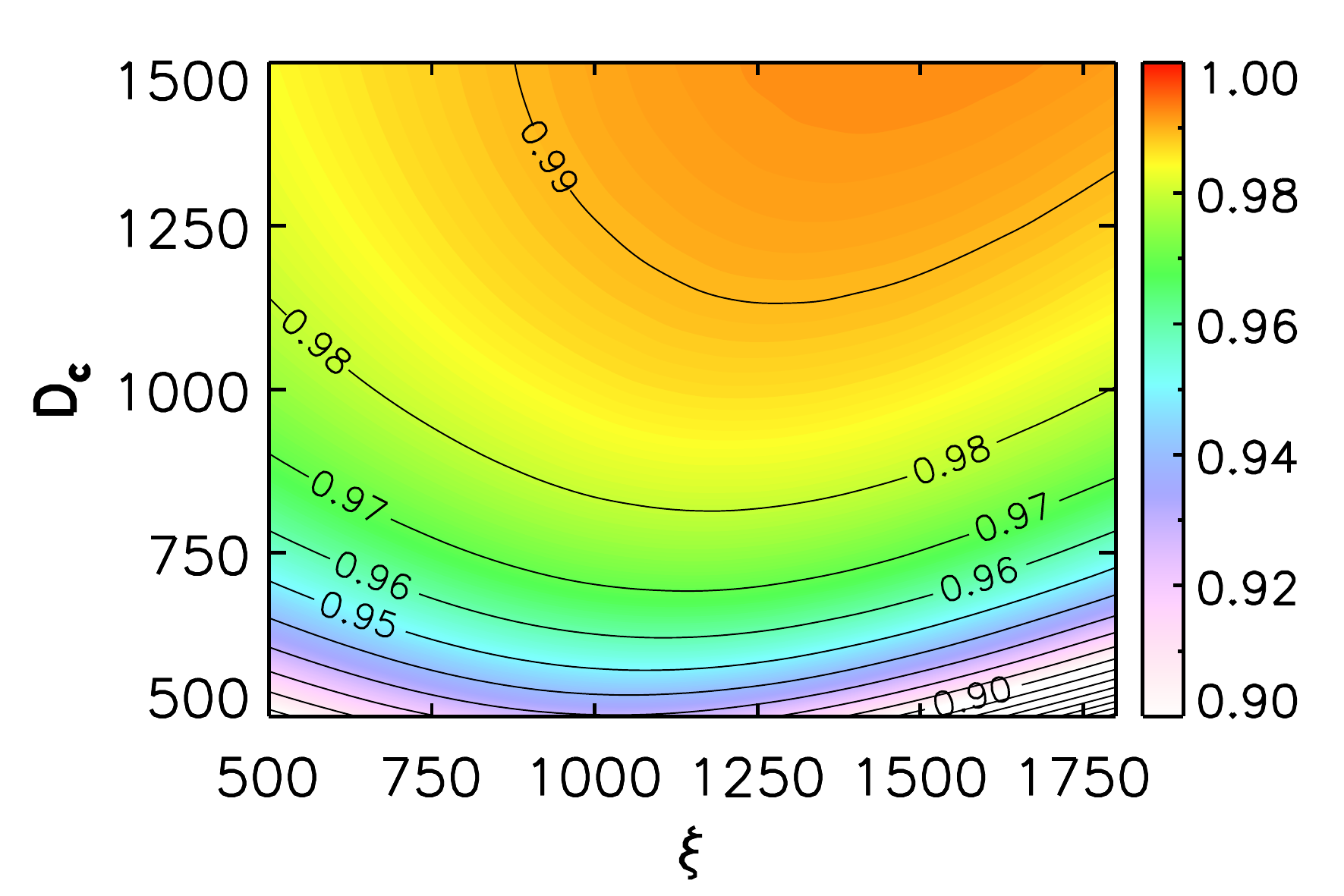}
  \caption{\label{fig3}
(Color online). The mean fidelity of the disordered echo memory with correlation length $\sigma=L/100$ of $\Omega_c$ at various optical depths $\xi$ and control field strengths $\mathbf{D}_\mathbf{c}$. The results are obtained by averaging over $1000$ realizations of different disorder.  
  }
\end{figure}
%

The value of encrypted data processing and storage has been proven by its benefits of protection  against data breaches and theft through the physical removal of computer memory \cite{Barker2015,Hellekalek2003}. An encrypted memory is composed of two components, namely, data storage and encryption.
Over past decades photon has been regarded as an ideal information carrier, and several schemes of optical quantum memory \cite{Lvovsky2009,Brennen2015} have been well established, e.g., electromagnetically induced transparency (EIT), \cite{Kocharovskaya1986,Boller1991,Hau1999,Kocharovskaya2001,Phillips2001,Chen2013,Cho2010,Hsiao2016}, 
gradient echo memory (GEM) \cite{Hetet2008b,Hosseini2009,Sparkes2010,Liao2014c,Chaneliere2015}, and
atomic frequency comb (AFC) \cite{deRiedmatten2008,Afzelius2009}. The missing piece of optical encryptor may open a vast  unexplored realm of   encrypted photonic memory at a security level higher than existing optical storage systems.
In this letter, we theoretically demonstrate not only the capability but also the  flexibility of using disordered fields \cite{Kalachev1998, Nefed2008,Billy2008} to set passwords on various schemes of photonic memory.
Incoporating spread spectrum methods \cite{Belthangady2010} with our scheme, long-distance quantum communication with high degree of multiplexing and noise immunity can be implemented via trusted and secure nodes. Our scheme may also pave the way to encrypt higher dimensional information, e.g., 2-dimensional images \cite{Shuker2008,Cho2012,Heinze2013,Ding2013,Lee2013}.

As illustrated in Fig.~\ref{fig1}, an echo memory \cite{Moiseev2001} is typically made of an atomic ensemble (gray spheres) whose absorption of a probe pulse (blue wavy lines) is modified by another external control field. The following phase of atomic coherence evolves in a nonuniform way due to the position-dependent control field strength. When control field is inverted at some instant, the time reversal of atomic phase leads to an echo output. 
In  Fig.~\ref{fig1}(a), we put forward the idea of disordered echo memory (DEM) using a disordered control field (coloured surface below gray spheres) whose strength statistics follows the normal distribution depicted in  Fig.~\ref{fig1}(c). 
Green arrows aligned aside the system represent the spatial randomness of DEM atomic dynamics.
As a comparison Fig.~\ref{fig1}(d) shows a linearly GEM having ordered atomic dynamics and equal probability for each field strength (Fig.~\ref{fig1}(e)).
We find that DEM leads to very high fidelity of broadband retrieval \cite{Reim2010,deRiedmatten2010}  when the strength statistics of the disordered field  (Fig.~\ref{fig1}(b)) matches the Fourier spectrum of incident probe (Fig.~\ref{fig1}(c)).
Besides, novel methods of disorder-induced symmetric encryption on optical memories will be presented.

Figure~\ref{fig2}(a) illustrates the DEM based on the three-level $\Lambda$-type scheme  depicted in Fig.~\ref{fig2}(b). 
The medium of length $L$ includes two ground states $\vert 1\rangle$, $\vert 2\rangle$, and an excited state $\vert 3\rangle$ whose lifetime is $\tau$ and spontaneous decay rate $\Gamma = 1/\tau$. 
The system under study is described by the optical-Bloch equation \cite{Scully2006,Hosseini2009,Chen2013,Liao2014c}
\begin{eqnarray}
 &  & \partial_{t}\hat{\rho}=\frac{1}{i\hbar}\left[\hat{H},\hat{\rho}\right]+\hat{\rho}_{dec},\nonumber\\
 &  & \left(\frac{1}{c}\partial_{t}+\partial_{z}\right)\Omega_{p}=i\eta\rho_{31},\nonumber
\end{eqnarray}
where $\hat{\rho}$ is the density matrix for the state vector
$\sum_{i=1}^{3}B_{i}\vert i\rangle$ of the 3-level atom, and each coherence $\rho_{ij}=B_i B_j^*$; $\hat{\rho}_{dec}$
describes the spontaneous decay of the excited state $\left|3\right\rangle $
characterized by rate $\Gamma$; $\hat{H}=\frac{\hbar}{2}\left[\Omega_{p}\left|3\right\rangle \left\langle 1\right|+\Omega_{c}\left|3\right\rangle \left\langle 2\right|+H.c.\right] $ is the Hamiltonian
describing the atom-light interaction with Rabi frequency $\Omega_p$  of the probe field and $\Omega_c$ of the control field; $\eta=\Gamma \xi/2L$ with $\xi$ and $L$ the optical depth and the length of
the medium, respectively. 
The boundary condition gives the Rabi frequency of the incident probe field $\Omega_p e^{-\left[ \left( t-t_p\right)/\kappa \right]^2 }$, and the initial condition describes the initial population $\rho_{11}(0,z)=1$,  $\rho_{ii}(0,z)=0$ for $i>1$ and zero coherences $\rho_{ij}(0,z)=0$ for $i\neq j$. 
Two dipole transitions $\vert 1\rangle\leftrightarrow \vert 3\rangle$ and $\vert 2\rangle\leftrightarrow \vert 3\rangle$ are resonantly coupled to a probe field 
and a disordered control field with Rabi frequency $\Omega_c(t,z)$, respectively.
The incident broadband probe field peaks at $t_p$ with a duration of $\kappa$ ($1/\kappa \gg \Gamma$), and the disordered control field satisfies the spatial correlation function $\left\langle \Omega_c\left(t,z\right)\Omega_c\left(t,z'\right)\right\rangle = \mathbf{D}_\mathbf{c}^2e^{-(z-z')^2/\sigma^2}$. 
The generation of disorder is given by Ref.~\cite{Simon2012}.
Here we consider  $\kappa=5\times 10^{-3}\tau$ and the correlation length $\sigma=L/100$.
The present broadband DEM relies on the spatially inhomogeneous Autler-Townes splitting (ATS) \cite{Nefed2008,Sparkes2010,Liao2012b,Liao2014c,Chaneliere2015} induced by $\Omega_c(t,z)$  such that each Fourier component of probe field is absorbed by the modified medium at correspondingly resonant positions \cite{Sparkes2010}. The probe field is then stored as atomic coherence $\rho_{21}(t,z) \propto  \sin\theta(t,z)$ and $\rho_{31}(t,z) \propto  \cos\theta(t,z)$
where the local phase is \cite{Liao2014c}
\begin{eqnarray}\label{eq1}
\theta(t,z)=\frac{1}{2}\int_0^{t}\Omega_c(t',z)dt'.
\end{eqnarray}
By inverting $\Omega_c(t,z)$ at some instant $t_\mathrm{i}$, i.e., $\Omega_c(t\geq t_\mathrm{i},z)= -\Omega_c(t<t_\mathrm{i},z)$,  an echo signal will be retrieved due to the global phase revival of $\theta(t,z)$.
For a broadband storage we use $\mathbf{D}_\mathbf{c}>1/\kappa \gg \Gamma$ which spreads absorption linewidth enough to cover the whole spectrum of the probe pulse. 
The success of the encryption is based on the cooperation between the disorder field and time reversal of phase $\theta(t,z)$.
As illustrated in Fig.~\ref{fig2}(a), we use a disordered control field depicted in Fig.~\ref{fig2}(e) with $\mathbf{D}_\mathbf{c}=1000\Gamma$, $t_\mathrm{i} = 0.22\tau$ and optical depth $\xi=1200$  to simultaneously store and encrypt the broadband probe (Green dashed line). The red solid line shows the only successful retrieved echo by using key1 profile demonstrated in Fig.~\ref{fig2}(f) where $\Omega_c(t\geq t_\mathrm{i},z)=-\Omega_c(t< t_\mathrm{i},z)$, namely, the inversion of the encryption key in Fig.~\ref{fig2}(e).
To test the security of the present encryption scheme, $\Omega_c(t\geq t_\mathrm{i},z)$ is replaced by a distinct disorder as key2 depicted in Fig.~\ref{fig2}(g) and a linearly gradient field as key3 in Fig.~\ref{fig2}(h) for decryption tests.
Blue long-dashed line and gray dotted line show that two attempts on memory both fail. 
Another 100 different disorders are also utilized to descrypt the echo, but none is successful.

The advantage of $\Lambda$-type DEM  is depicted in Fig.~\ref{fig3}, i.e., a high fidelity 
\begin{equation}
\mathcal{F} = \frac{\vert \int_{t_\mathrm{i}}^\infty\Omega_p^*(t-t_d,0)\Omega_p(t,L) dt\vert^2}{\left[ \int_{t_\mathrm{i}}^\infty\vert\Omega_p(t,0)\vert^2 dt\right] \left[\int_{t_\mathrm{i}}^\infty\vert\Omega_p(t,L)\vert^2 dt\right]}\nonumber
\end{equation}
approaching unity for a wide range of $\mathbf{D}_\mathbf{c}$ and $\xi$. 
Also, we numerically  observe that in the domain of interest the present broadband scheme may share the same upper limit of storage efficiency  
$\mathrm{SE} =\left[  \int_{t_\mathrm{i}}^\infty\vert \Omega_p(t,L) \vert^2 dt\right]  / \left[ \int_{-\infty}^\infty\vert \Omega_p(t,0) \vert^2 dt \right] 
$
of around $54 \%$ with other echo memory schemes because of reabsorption \cite{Hosseini2009,Afzelius2009}. This limit can be overcame by the cavity-assisted interaction \cite{Sabooni2013,Jobez2014}.
The underlying reason for the very high fidelity is that
the solution of an echo signal can be replaced by the inverse Fourier transformation 
$\int_{-\infty}^\infty\mathrm{P(\Omega_c)}\cos\left[ \frac{\Omega_c  (t-t_\mathrm{i}) }{2}\right] d\Omega_c$. The  statistics $\mathrm{P(\Omega_c)}$ behind disordered field strength preserves the shape of the incident probe field when $\mathrm{P(\Omega_c)}\approx \int_{-\infty}^\infty\Omega_p(t,0)\cos\left[ \frac{\Omega_c  t }{2}\right] dt$. Our numerical result in Fig.~\ref{fig3} confirms this optimization of fidelity when $\mathbf{D}_\mathbf{c}\geq 1/\kappa$, i.e., the  width of the absorption spectrum of the memory greater than or equal to
that of  the Fourier transformation of the incident pulse.

To demonstrate the encryption on a narrowband ($1/\kappa < \Gamma$) EIT-based memory, we 
invoke the time sequence and the  four-level $N$-type system respectively illustrated in Fig.~\ref{fig2}(c) and (d).
In Fig.~\ref{fig2}(c), the green dashed line shows the incident probe pulse with $\kappa=10\tau$ driving the transition $\vert 1\rangle\leftrightarrow \vert 3\rangle$.
The brown dashed dotted line illustrates the temporal sequence of the spatially uniform control field with Rabi frequency $\Omega_c =5\Gamma$, which  drives transition $\vert 2\rangle\leftrightarrow \vert 3\rangle$. The EIT light storage is accomplished by turning off control field at $t_\mathrm{off}=50\tau$, and subsequently on at $t_\mathrm{on}=120\tau$ for light retrieval \cite{Fleischhauer2000,Kocharovskaya2001,Liu2001,Phillips2001,Chen2013,Hsiao2016}.
As demonstrated in Fig.~\ref{fig2}(d), during the storage period $t_\mathrm{off}<t<t_\mathrm{on}$, a third switching field with Rabi frequency $\Omega_s (t,z)$ shines on the memory to drive transition $\vert 2\rangle\leftrightarrow \vert 4\rangle$.
The $\Omega_s (t,z)$-encrypted atomic excitation is given by
$
\rho_{21}(t,z) \propto \cos\phi(t,z)
$
and 
$
\rho_{41}(t,z) \propto i\sin\phi(t,z)
$
where the local phase is
\begin{equation}\label{eq2}
\phi(t,z)=\frac{1}{2}\int_{t_\mathrm{off}}^{t>t_\mathrm{off}}\Omega_s(t',z)dt'.
\end{equation}
It is the local phase modulation making encryption possible such that one can locally switch the excitation between $\rho_{21}(t,z)$ and $\rho_{41}(t,z)$ by engineering the spatial profile of $\Omega_s (t,z)$.
In Fig.~\ref{fig2}(c) with $\xi=1200$, orange dashed-dotted line shows that  two spatially disordered  pulses $\Omega_s (t,z)$  encrypt and decrypt the memory with keys respectively in Fig.~\ref{fig2}(e) and Fig.~\ref{fig2}(f).
Two disorders lead to $\phi(t\geq 110\tau,z)=0$ and obey $\left\langle \Omega_s\left(t,z\right)\Omega_s\left(t,z'\right)\right\rangle =\mathbf{D}_\mathbf{s}^2e^{-(z-z')^2/\sigma^2}$ where $\mathbf{D}_\mathbf{s}=30\Gamma$ and $\sigma=L/100$. To also check whether other keys can recover the encrypted memory, we use key2 and key3 for decryption. As demonstrated in Fig.~\ref{fig2}(c), 
blue long-dashed line and gray dotted line reveal that
no signal is retrieved by using keys other than key1 when $\Omega_c$ is turned on. The retrieval is also futile when either encryption pulse $(60\tau <t< 80\tau)$  or decryption pulse $(90\tau <t< 110\tau)$ is applied.
Moreover, the decrypted signal peaking at 150$\tau$ in Fig.~\ref{fig2}(c) perfectly matches the pure EIT retrieval without any phase modulation, namely, the disordered encryption procedure preserves the performance of an EIT-based memory.

\begin{figure}[t]
\vspace{-0.4cm}
  \includegraphics[width=0.45\textwidth]{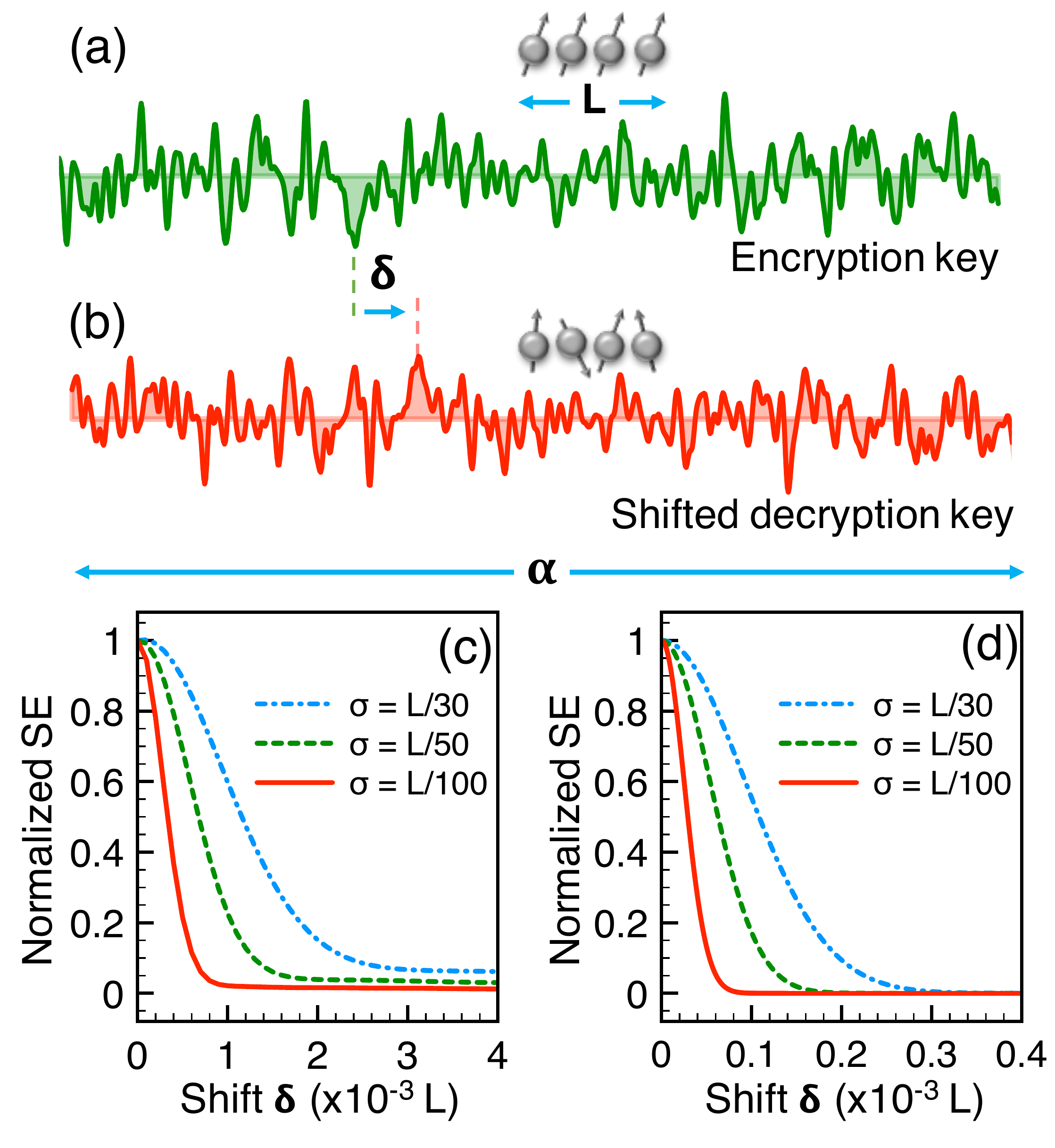}
  \caption{\label{fig4}
(Color online).
(a) Disordered encryption key with length $\alpha$ interacting with a memory of size $L$. 
(b) Decryption key is made of the inversion of encryption key spatially shifted by $\delta$.
The mean $\delta$-dependant normalized storage efficiency over 100 realizations for 
(c) disordered echo memory and
(d)  EIT-based memory.
$\mathbf{D}_\mathbf{c}=1500\Gamma$ and $\xi=1500$ are used in both (c) and (d).
Three lines correspond to different correlation lengths of disorder $\sigma=L/100$ (red solid),  $\sigma=L/50$ (green dashed) and $\sigma=L/30$ (blue dashed dotted).
  }
\end{figure}

To verify the degree of the confidentiality of our scheme, we consider the case that the stored optical memory is retrieved by not only inverting the encryption key but also with a spatial shift $\delta$ as depicted in Fig.~\ref{fig4}(a) and (b). At $\delta=0$, the mean storage efficiency averaged over $100$ realizations is maximized as expected, as increasing $\delta$ the storage efficiency gradually decreases as shown in Fig.~\ref{fig4}(c) and (d) for $\Lambda$-type and $N$-type schemes, respectively. Those faint retrievals in  Fig.~\ref{fig4}(c) with normalized SE of about 2\%-5\% at large $\delta$ for $\sigma = L/50$ and $\sigma = L/30$ are too noisy to determine their fidelity.
The very narrow  $\sigma$-dependant window of retrieval ($< 10^{-3} L$ for $\Lambda$-type DEM and  $< 10^{-4} L$ for EIT-based memory) suggests a securer encryption scheme that
one can prepare a much longer disorder with length $\alpha$ than the memory size $L$, but only a tiny and privately known section is applied. 
In order to access the stored photonic data, the precise knowledge of the used section is needed, which assure the confidentiality. 
The degree of the confidentiality of the scheme can be defined as 
$\chi = \frac{\alpha^2}{\sigma L}$,    
which reflects the increase of confidentiality by  using  the condition $\alpha\gg L \gg\sigma$.
A study of the error bars of the averaged fidelity of  $\Lambda$-type DEM also suggests that
a small $\sigma$  guarantees both security and  identical retrievals from one disorder to another for a given incident probe pulse. 

To raise the security level of encryption, a pair of keys can alternatively be a correlation function and a set of random numbers, which constitute a disorder \cite{Simon2012}. 
Such an implementation is resistant against brute-force attack if the key length of the pseudorandom number generator (PRNG) is large enough. More tenable attacks would come in the form of cryptanalysis, where the adversary exploits the system vulnerabilities either directly, through the inputs, or the internal states of the PRNG. Approaches to enhance the cryptographic strength of the PRGN against such attacks employ cryptographic primitives like the hash function; and by gathering randomness from external sources to maximise its entropy \cite{Menezes1997}. Cryptographically secure PRNG passes a suite of statistical tests \cite{Rukhin2010}, with good examples being Advanced Encryption Standard \cite{Hellekalek2003} and the PRNGs published by National Institute of Standards and Technology \cite{Barker2015}.
In conclusion, we demonstrate using disordered fields to encrypt  photonic memories with both high fidelity and high confidentiality. We expect that the present method  is useful for not only the longitudinal but also the transverse encryption, and the latter can randomize and recover the phase of the stored images \cite{Shuker2008,Cho2012,Heinze2013,Ding2013,Lee2013}.  
Our scheme significantly lowers the intrinsic risk of data loss in existing memory schemes
and makes an impact on photonic information technology.

S.-W.~S. and S.-C.~G. are supported by the Ministry of Science and Technology, Taiwan (Grant No. MOST 103-2112-M-018-002-MY3). 
A. K. acknowledges the support from the Russian Science Foundation (Grant No. 14-12-00806).
W.-T.~L. is supported by the Ministry of Science and Technology, Taiwan (Grant No. MOST 105-2112-M-008-001-MY3). 
S.-C.~G. and W.-T.~L. are also supported by the National Center for Theoretical Sciences, Taiwan. 


\bibliographystyle{apsrev}
\bibliography{echo}

\end{document}